\documentstyle[psfig]{laa}
\begin{document}
\thesaurus{08.01.1 10.08.1 12.03.3}

\title{Chemical evolution of Damped Ly$\alpha$ systems}

\author{ F. Matteucci \inst{1},  P. Molaro \inst{2}, 
 G. Vladilo  \inst{2} }

\offprints{F. Matteucci, 
 Dipartimento di Astronomia, Universit\`a di Trieste,
SISSA, Via Beirut 2-4, I-34013 Trieste, Italy
}

\institute{
        Dipartimento di Astronomia, Universit\`a di Trieste,
SISSA, Via Beirut 2-4, I-34013 Trieste, Italy
\and
        Osservatorio Astronomico di Trieste, Via G.B. Tiepolo 11,
        I-34131 Trieste, Italy 
}

\date{Received date; accepted date}
\maketitle

\begin{abstract}

High redshift DLA systems suggest that the 
relative abundances of elements might be roughly  solar, although
with absolute abundances of more than two orders of magnitude below
solar.
The result comes  from observations of the [SII/ZnII] ratio, which is 
a reliable diagnostic of the true abundance, and from DLA absorbers with 
small 
dust depletion  and negligible HII contamination. In particular,
in two DLA systems nitrogen is detected and at remarkably  high levels
(Vladilo et al. 1995,
Molaro et al. 1995, Green et al. 1995, Kulkarni et al. 1996).
Here we compare the predictions from chemical evolution models of
galaxies of different morphological type with the abundances and
abundance ratios derived for such systems. 
We conclude
that solar ratios and relatively high nitrogen
abundances
can be obtained  in the framework
of a chemical evolution model assuming short but intense
bursts of star formation, which in turn trigger enriched galactic
winds, and a primary origin for nitrogen
in massive stars. Such a model
is the most successful
in describing the chemical abundances of dwarf irregular galaxies and in 
particular of the peculiar galaxy IZw18.
Thus,  solar ratios at very low absolute abundances, if confirmed,
seem to favour 
dwarf galaxies rather than spirals as the progenitors of at least 
some of the DLA systems.

\end{abstract}

\section{Introduction}

Absorption line systems detected in the spectra of Quasi Stellar Objects
(QSO) originate in 
intervening galaxies or protogalaxies. 
Among the different classes of absorbers, the damped Lyman $\alpha$  
systems (DLA) 
are those characterized by N(HI) $\ge$  10$^{20}$ cm$^{-2}$
and by showing many low ionization species such as FeII, SiII, CrII, ZnII 
and OI, which are 
in their dominant ionization stage for a HI region.
The high accuracy in the hydrogen column density determination
derived from the damped wings 
and the absence of significant ionization 
corrections allow accurate absolute 
abundance determinations of the gas
phase elements. Since 
DLA  systems
are observable up to the highest redshift they provide a unique tool for the
study of  the cosmic chemical history of the early universe. 
Such observations are complementary to the Hubble Deep Field 
ones (Mobasher et al. 1996) which show that high redshift galaxies 
have a large variety of 
morphological types including many peculiar objects with traces of 
interaction and merging: interestingly, a large fraction are starburst 
galaxies, 
as suggested by their colors.

The relative elemental abundances rather than absolute ones
are a valuable diagnostic of the first elemental buildup.
This is because absolute abundances are affected by all the
model assumptions, whereas abundance ratios generally depend
only on the assumed nucleosynthesis and stellar lifetimes. 
The way of using the information contained in abundance ratios 
is to compare the observed ratios with predictions from chemical evolution models which
take into account detailed stellar nucleosynthesis and lifetimes.
As in the Milky Way, we hope to distinguish halo-like abundances,
produced by type II SNe and characterized by enhancements of $\alpha$
elements with respect to the iron-peak elements, from disk-like abundances,
produced by the cumulative effect of both type Ia and type II SNe, where the 
abundance ratios become progressively solar.
The halo abundance pattern follows from the  fact that 
Fe arises mostly 
from type Ia supernovae with some contribution from type II,
while the reverse is true for Si and O. The longer lifetimes of the SNIa, 
which are believed to have progenitors with masses of
1-8 M$_{\odot}$, result in a delayed iron
enrichment compared to the major SNII products such as $\alpha$-elements
(Tinsley, 1980; 
Greggio and Renzini, 1983a;
Matteucci and Greggio, 1986). Therefore, 
the  behaviour of the [$\alpha$/Fe] ratios is
mainly dependent on the
relative lifetimes of type Ia and II supernovae.  
Another key element is nitrogen and the ratio N/O. 
This element, in fact, is thought to originate mainly from low
and intermediate mass stars and to be a "secondary" element,
in the sense that it is produced proportionally to the initial
stellar metallicity. As a consequence nitrogen is restored into
the interstellar medium with a large temporal delay relative to
oxygen, which is produced in massive stars and is a "primary" element,
namely its production is independent of the initial stellar metallicity.
For this reason, the $\alpha$/Fe and N/O ratios
can be used as cosmic clocks and
represent a clue in understanding the nature 
of high redshift objects such as QSO (Hamman and Ferland, 1993; 
Matteucci and Padovani, 1993) and DLA systems (Matteucci, 1995).

The DLA systems are generally believed to be the progenitors
of the present-day spiral galaxies (Wolfe et al. 1986). However, it 
has also been suggested that they could be dwarf galaxies, 
since they have a large amount of gas, low metallicities
and low dust-to-gas ratios (Pettini et al. 1990;Meyer and York 1992;
Steidel 1994).
Whether DLA are proto-spirals or proto-dwarfs can be indicated by
their observed abundance pattern. To this purpose
 in this paper we will present models of chemical evolution for galaxies
of different morphological type and compare them with the observed abundances
in DLA.

\section{Observed abundances in damped Ly$\alpha$ systems} 

So far chemical  abundances have been measured in a set of DLA systems 
and with a variety of resolutions and accuracies by
Black et al. (1987),
Meyer \& York (1987),
Chaffee et al. (1988),
Meyer et al. (1989), 
Meyer \& Roth (1990),
Rauch et al. (1990),
Pettini et al. (1990),
Meyer \& York (1992),
Pettini et al. (1994),
Fan \& Tytler (1994),
Wolfe et al. (1994),
Pettini et al. (1995),
Lu et al. (1995a),
Steidel et al. (1995), Lu et al (1995b). Here,
we would like to focus on the abundances which could be used
to compare with  
chemical evolution 
models and in particular with 
the few  nitrogen detections so far reported in the literature.
A summary  of the abundances for the systems for which nitrogen detection or
upper limits are available are reported in Table 1 and 2.
All the original abundances have been renormalized to 
the solar abundances of Anders \& Grevesse (1989), with the exception of
$\log (Fe/H)_{\sun}$=$-4.52$ taken from Hannaford et al. (1992) 
([M/H] $\equiv$ log(M/H)-log(M/H)$_{\sun}$). 
The abundances in the Tables are
derived from column density ratios of different atomic and ionic species
in the gas phase,  and 
can  be translated to elemental abundances only if ionization effects 
and dust depletion are negligible
or can be accounted for. In the following we discuss shortly these two
effects.

\begin{table*}
\caption{ 
Hydrogen column densities and elemental
abundances in DLA systems with NI measured. 
Original abundances have been renormalized to the solar values 
log (N/H)$_{\sun}$ = $-$3.95, log (O/H)$_{\sun}$ = $-$3.07, 
log (Si/H)$_{\sun}$ = $-$4.45,
log (S/H)$_{\sun}$ = $-$4.73, taken from Anders \& Grevesse (1989),
and log (Fe/H)$_{\sun}$ = $-$4.52 from Hannaford et al. (1992). 
}
\begin{flushleft}
\begin{tabular}{lclllllll}
\hline 
~~~QSO    & $z_{\rm abs}$ & N(HI)  & ~~~[N/H] & ~~~[O/H] & ~~~[Si/H]   
          &  ~~~[S/H] &   ~~~[Fe/H] & Refs. \\
\hline \hline
          &&&&&&&\\
0000-26   & 3.3901 & 21.30 & $-2.77^{+0.17}_{-0.17}$ & $-3.13^{+0.17}_{-0.17}$ 
          & $-2.48^{+0.19}_{-0.19}$ 
          & ~~~~---   & $-2.38^{+0.16}_{-0.16}$ &   ~~1    \\ 
          &&&&&&&\\
1331+1700 & 1.7765 & 21.18 & $-2.73^{+0.11}_{-0.11}$ & $-2.81^{+0.21}_{-0.30}$ 
          & ~~~~--- & $-1.35^{+0.11}_{-0.11}$ 
          & $-2.25^{+0.11}_{-0.12}$ & ~~2$^a$ \\ 
          &&&&&&&\\
2348-147  & 2.2794 & 20.57 & $<-3.15$ & $-2.13^{+2.26}_{-0.58}$ 
          & $-1.97^{+0.13}_{-0.10}$ & $-1.91^{+0.14}_{-0.16}$ 
          & $-2.35^{+0.22}_{-0.10}$ & ~~3 \\
          &&&&&&&\\
2344+124  & 2.5379 & 20.43 & $-3.00^{+0.12}_{-0.22}$ $^b$ 
          & $-2.11^{+2.5}_{-0.36}$ 
          & $-1.72^{+0.27}_{-0.13}$ & $<-1.20$  
          & $-1.85^c$ & ~~4 \\ 
          &&&&&&&\\
1946+7658 & 2.8443 & 20.30 & $<-3.26$ $^d$ & $-2.65^{+0.15}_{-0.15}$  
          & $-2.16^{+0.10}_{-0.10}$
          & $<-0.89$ & $-2.41^{+0.10}_{-0.10}$ & ~~5$^e$ \\
          &&&&&&&\\
\hline 
\end{tabular}
\end{flushleft}

\vskip 0.4 truecm

\begin{flushleft}
 
References for Table 1 \\

(1) Molaro et al. (1995) ; (2) Green et al. (1995) ; 
(3) Pettini et al.(1995); \\
(4) Lipman, thesis (1995) ; (5) Lu et al. (1995a)

\end{flushleft}

\vskip 0.4 truecm

\begin{flushleft}

$a$ : Hydrogen column density from Pettini et al. (1994);
       metal abundance error bars
       include propagation of the N(HI) error. 

$b$ : Nitrogen value derived by using only 
       the main absorption component of the system;
     
$c$ :  Error bar not reported by the author.

$d$ : Upper limit from Lipman (1995; thesis). 

$e$ : Column densities from line profile fitting where all the
      lines are fitted simultaneously; the final choice of Lu et al (1995a)
for O is [O/H]$>$ -2.99; metal abundance 
      error bars include propagation of the N(HI) error.       

\end{flushleft}
\end{table*}

\begin{table*}
\caption{Element to element abundances for the DLA systems
listed in Table 1.}
\begin{flushleft}
\begin{tabular}{lcrrrrrrr}
\hline 
~~~QSO  & $z_{\rm abs}$ & ~~~[N/O] & ~~~[N/Si] & ~~~[N/S]& ~~~[Si/O] & ~~~[S/O] & ~~~[O/Fe] &
        ~~~[Si/Fe] \\
\hline
\hline
         &&&&&&&\\
0000-26  & 3.3901  &  +0.36   & $-$0.29   & ---~ & +0.65  & ---~ & $-$0.75 &
         $-0.10$\\ 
         &&&&&&&\\
1331+1700 & 1.7765 &  +0.08     &  ---~ &  $-$1.38 & ---~  & +1.46 & $-$0.56 &
         ---~   \\ 
         &&&&&&&\\
2348-147 & 2.2794  &  $<-1.02$  & $<-1.18$ & $<-1.24$ & +0.16 & +0.22 & +0.22 &
         +0.38  \\ 
         &&&&&&&\\
2344+124 $^a$ & 2.5379
         & $-0.59$ & $-1.14$ & $>-1.32$ & +0.55 & $<+0.73$ & $-0.59$ &
         +0.13  \\ 
         &&&&&&&\\
1946+7658 & 2.8443 & $<-0.61$ & $<-1.10$ & ---~ & +0.49 & $<+1.76$  &  $-0.24$ &
         +0.25  \\
         &&&&&&&\\
\hline 
\end{tabular}

\vskip 0.5 truecm

$a$ Values derived by using only the main absorption component

\end{flushleft}
\end{table*}

\subsection{Ionization corrections and dust depletions}

The neutral hydrogen in the DLA is optically thick to the ionizing radiation
either from the intergalactic background or from starlight inside the DLA,
and the abundances derived from species that 
have ionization potentials in excess of 13.6 eV 
do not require ionization corrections. 
This is the case of abundances derived from 
NI, OI, AlII, SiII and FeII, which 
are dominant ionization stages in the HI gas in our own Galaxy.
Detailed ionization models for intergalactig radiation field at high redshift explored by Lu et al (1995) and Fan and Tytler
(1994) show that ionization  corrections are indeed minimal for 
systems with $\log$ N(HI)$>$ 20.

Another effect which could affect the abundance determinations 
is the presence of HII regions within the DLA  intercepted by 
the line of sight
(Steigman et al 1975). These regions of warm ($T$ $\simeq$ 10$^4$ K)
ionized gas would contribute
to the column densities of singly ionized species, such as FeII, SiII
and AlII, but not to the HI column density.
In the interstellar gas of our Galaxy the column density 
contribution of HII regions is generally negligible 
when the HI column density is as high as it is in DLA systems
(N(HI) $\ga$ 10$^{20}$ cm$^{-2}$). 
If HII regions were more important in DLA than in our Galaxy they would 
produce different line profiles between neutral and 
ionized lines, when observed at high resolution.
This effect is not observed in  the DLA towards QSO 0000-2169 where the $b$ values, velocities and profiles of singly ionized species are
the same as those of the neutral species, suggesting that both form
in the same slab of HI material. 
Extra contributions to the absorption from HII regions in this DLA
are also constrained 
by the lack of detection of NII, resulting in NII/NI $<$ 0.4. 
Viceversa, 
Green et al. (1995) interpret the abundances of the DLA  
towards MC3 1331+170 
invoking an extraordinary amount of ionized gas, namely six times the
neutral gas, which makes rather uncertain the abundances of ionized species.

Highly ionized gas in the form of SiIV and CIV is frequently observed
in the DLA. Sometimes the highly ionized gas is found at the same velocities
of the neutral one  but often is found
at different velocities and therefore is very probable that highly
ionized gas  arises either  in regions 
disconnected from the neutral material, or in the interfaces between
neutral material and the intergalactic medium.

Presence of dust will selectively deplete the elements 
observed in the gas phase.
In the dense clouds in the Galaxy the interstellar depletion of Al
and Fe can reach  $\simeq$ 2 dex, that of Si $\approx$ 1 dex, while, 
on the other extreme the  O depletion 
is $\le$ 0.4 dex, that of N   
$\le$ 0.2 dex and S essentially undepleted (Jenkins 1987). 
Reddening measurements of QSO with DLA systems suggest  
a dust-to-gas ratio in the DLA
of about 10\% of that in our Galaxy (Fall and Pei 1989). 
In the survey of ZnII and CrII in DLA by  Pettini et al.  (1994)
chromium is typically one dex below zinc, while in the Galaxy 
is about two dex, showing the presence of dust and  
a reduced dust depletion 
in the DLA systems. 
If some dust were present  
this would alter the abundance determinations of refractory 
elements, such as   
Fe, Si and Al.
Much more reliable  are  the
abundances 
of the non-refractory elements, such as
Zn, S, O and N.

\subsection{$\alpha$ versus iron-peak elements}

Of particular importance for understanding galactic chemical evolution
is the comparison between $\alpha$ 
and iron peak elements. 
In particular, the ratio between sulphur and zinc is
an important diagnostic tool.
Both elements
show little affinity with dust and 
their ratio is also safe against  possible contributions
from HII regions, since these essentially cancel out. 
Considering that S and Zn 
have different nucleosynthetic origin with S mainly a product of type II SN
and Zn of type Ia SN (Matteucci et al. 1993),  
the [SII/ZnII] ratio is an ideal diagnostic 
for understanding the character of the chemical evolution. 
So far, only few determinations of
[S/Zn] ratio in DLA are present in the literature.
Meyer et al. (1989) found [S/Zn]=$-$0.1
in the DLA at $z=2.8$ towards 
QSO PKS 0528-250 and Green et al. (1995) found [S/Zn] = +0.1 
in the $z$=1.775 DLA in Q1331. Thus,
in spite of the very poor metallicity of the gas, 
which in these systems is $\simeq$ -2.0, the  ratios of elements
non depleted in dust  have solar values.

In DLA systems 
the relative abundances of Si and Fe are often found
to be consistent with a halo pattern.
Lu et al. (1995a) found 
[Si/Fe]=+0.31 for the DLA at $z=2.844$ towards 
HS 1946=76 and Pettini et al.
found [Si/Fe]=+0.4 for the DLA at $z=2.27$ towards QSO 2348-147. 
However, Si and Fe are differentially depleted from gas to dust
in our Galaxy. The average value in the compilation of 11 clouds observed
with HST made by Lu et al (1995a, their Table 8) is 
[Si/Fe]=+0.66$\pm0.26$.
The observed enhancement of Si versus Fe in the DLA would 
reflect the overabundances
of $\alpha$ elements with respect to the iron-peak elements only 
in complete absence of dust.
Since DLA show some amount of dust, 
a moderate enhancement of Si over Fe cannot be considered as a
clear-cut evidence of a halo-like pattern. 
Moreover, values of [Si/Fe]$\ge$0.6, that would be
expected if differential depletion and intrinsic $\alpha$
enhancement are present, have not yet been observed.

\subsection{Nitrogen and oxygen abundances}

Nitrogen has been detected 
in the DLA at z=3.39 towards QSO 0000-26,  with  
[N/H] = $-$2.77 $\pm 0.17$ (Vladilo et al. 1995 and Molaro et al. 1995) and
in the DLA at $z$=1.78 towards MC3 1331+170 with 
[N/H]=-2.7 (Green et al. 1995).  
A significant
upper limit has been derived by Pettini et al. (1995) 
in the DLA systems towards
QSO 2348-147 NI$<$-3.15.
By means of co-addition technique of the lines of the 1200 \AA\ triplet
Lipman (1995) achieved a marginal detection of NI  
for the main component  of the
DLA towards QSO 2344+124.
 By using the data published by Fan and Tytler (1994), 
Lipman derived also an upper limit of [N/H] $<$ -3.26. 
These measurements   show  that a real dispersion in the nitrogen abundances 
may be present among the DLAs. 
 
In order to
understand the nucleosynthetic origin of nitrogen the ideal would be to
follow its abundance with respect to that of oxygen.
Unfortunately, 
oxygen abundance is generally given with a large uncertainty.
This is because oxygen abundance is derived from the OI 1302.1685  \AA\ line which
has log~gf=1.804 and is generally saturated.  The
 other line OI 1355.5977, has log~gf=-2.772 and is 
generally too faint to place useful upper limits.
The line saturation leaves the line broadening poorly constrained
and leads to the large uncertainty in the oxygen abundance.
However, the $b$ value can be inferred 
from other lines under the assumption that they 
are formed in the same material and that the origin of
broadening is not thermal.
Lu et al. (1995a) and Molaro et al. (1995) use the $\it b$ value 
from the other observed species thus removing   
the large  uncertainty in the
oxygen abundance (cfr. Table 1).
In Molaro et al. (1995). the broadening value is taken not only 
from ionized species, which might be sensitive 
to extra contribution from HII gas along the line of sight, but also from NI 
which is forming in the same material as neutral oxygen. 
It is rather striking that in the  cases with oxygen abundances with 
small associated errors,
oxygen turns out to be remarkably  deficient relatively
to the other elements measured, 
such as Si, S and Fe as it is possible to see in Table 2.

To circumvent the problem of oxygen uncertainty, 
Pettini et al. (1995) and Lipman (1995) take Si or S as a proxy for oxygen
assuming  [O/Si] = [O/S]=0. This assumption is about true for the halo
stars of our own Galaxy (however see later for silicon), 
but is rather risky for primeval galaxies
which might have experienced a different chemical evolution from that
of our own Galaxy. 
In particular, Si and O do not have the same nucleosynthetic
origin, silicon  may be  affected by dust depletion and
SII and SiII may be also affected by HII contribution.
Also by adopting the Pettini et al.  approach for  the case of QSO 0000-2169 
we have  [N/Si] = $-0.3$, which stands 
genuinely high when compared to the other determinations.

\section {What chemical evolution?}

Some DLA seem to show similarities to
dwarf irregular and blue compact galaxies 
in terms of abundance pattern.
These galaxies, in fact, 
exhibit relatively high N/O, although with a large spread,
at a low overall metallicity Z ranging from 1/10 to 1/30 of the solar value
(see Matteucci, 1995).
In particular, IZw18 is the galaxy with the lowest metal content known locally,
and a N/O ratio roughly solar.
Recently, a single starburst chemical evolution model has been
successfully developed by Kunth et al. (1995) to reproduce the
observed abundances of IZw 18.
Kunth et al. (1995) applied Marconi's et al. (1994) model where
star formation is assumed to proceed in short but intense bursts.
The contribution to the chemical enrichment of SNe
of different type (II, Ia and Ib) is taken into account.
The novel feature is the introduction of differential galactic winds
where some 
elements are preferentially lost from the galaxy relative to others.
Recently,
more and more observational evidence is growing about the
existence of such galactic winds in dwarf
irregulars as provided by 
Meurer et al. (1992), Lequeux et al.  (1995) and Papaderos et al.  (1994).
These winds are found to be a crucial ingredient 
in the model of Kunth et al. (1995) in order to reproduce
the high observed N/O ratio in IZw18.
To explain the nitrogen abundance of IZw18 this same model 
assumes also a
primary N production from massive stars, as described in the next section.
\par
In this paper we will show the predictions
of models similar to that of Kunth et al.
together with the predictions of models for the
chemical evolution of the solar region, under 
different assumptions about the production and nature of 
nitrogen, and we will
compare them with the DLA data.

\section{The chemical evolution model for dwarf irregulars}

We are using here the same model adopted by Kunth et al. (1995)
which is aimed at describing the evolution of IZw18, namely of an object
presently
having a strong burst of star formation which induces a galactic wind.

The main features of this chemical evolution model are the following: 

\par
1) one-zone, with instantaneous and complete mixing of gas inside this zone,
\par
2) no instantaneous recycling approximation; i.e. the stellar 
lifetimes are taken into account,
\par
3) only one intense burst of star formation is assumed to occur,
\par
4) the evolution of several chemical elements (He, C, N, O, Fe) due 
to stellar nucleosynthesis, stellar mass ejection, galactic wind 
powered by SNe and infall of primordial gas, is followed in detail.

\medskip
If G$_i$ is the fractional mass of the element i, its evolution is 
given by the equations described in Marconi et al. (1994).
$$
\dot G_i(t) = -\psi(t)(1+w_{i}) X_i(t)+ $$
$$ \int_{M_{L}}^{M_{Bm}}{\psi(t-\tau_m) 
Q_{mi}(t-\tau_m)\phi(m)dm}+  $$
$$ A\int_{M_{Bm}}^{M_{BM}}{\phi(m)}\bigl[\int_{\mu_{min}}
^{0.5}{f(\mu)\psi(t-\tau_{m1}) 
Q_{mi}(t-\tau_{m2})d\mu\bigr] dm}+ $$
$$ (1-A)\int_{M_{Bm}}^
{M_{BM}}{\psi(t-\tau_{m})Q_{mi}(t-\tau_m)\phi(m)dm}+$$   
$$\int_{M_{BM}}^{M_U}{\psi(t-\tau_m)Q_{mi}(t-\tau_m)+ 
\phi(m)dm} + \dot  G_{i}(t)_{inf} 
\eqno(1) $$
where $G_i(t)$=$M_{gas}(t)X_i(t)/M_{tot}(t_{G})$ is the mass 
density of gas in the form of an element {\it i} normalized 
to the total mass 
at the present time $t_{G}$.
The quantity $X_i(t) = G_i(t)/G(t)$ represents 
the abundance by mass of the element $i$ 
and by definition the summation over all the abundances
of the elements present in the gas 
mixture is equal to unity. 
The quantity $G(t) = M_{gas}/M_{tot}$ is the total 
fractionary mass of gas, 
and $M_{tot}$ refers 
only to the mass present in the form of gas in the star 
forming region.
The possible presence of a dark matter halo is not considered here, given
the simple treatment of the development of a galactic wind, as we will see
in the following.
\par
The star formation rate we assume during the burst, $\psi(t)$, is defined as:
$$
\psi(t)\,=-\nu \, \eta(t) \, G(t) 
$$

where $\nu$ is the star formation efficiency (expressed in units of  
Gyr$^{-1}$), and represents the inverse of the timescale of star 
formation, 
namely the timescale necessary to consume all the gas in the star 
forming region; $\eta(t)$ takes into account the 
stochastic nature of the star formation processes as in Gerola et al. 
(1980) and Matteucci and Tosi (1985),  
where a detailed description can be found.
 
The galactic wind is assumed to be simply proportional to the star formation
rate (namely to the rate of explosion of type II SNe). This is a reasonable 
choice since the duration of the burst is so short that it does 
not allow the explosion of type Ia SNe, which occur all after the burst. 
The simplicity of 
the treatment of the galactic wind
avoids also the introduction of other unknown 
parameters such as the efficiency of
energy tranfer from stars and supernovae to the ISM.
The rate of mass loss via a galactic wind is defined
as follows:\par
$$
\dot G_{iw}(t)\,=\, -w_{i}\psi(t)X_{iw}(t) 
$$
where $X_{iw}(t)=X_{i}(t)$ and $w_{i}$ is a free 
parameter containing all the 
information about the energy released by SNe and the efficiency with 
which such
energy is transformed into gas escape velocity 
(note that Pilyugin (1992;1993)
defines a wind parameter which is the inverse of $w_{i}$, namely the ratio
of the star formation rate to the wind rate).
The value of w$_{i}$ has been assumed to be 
different for different elements. 
In particular, the 
assumption has been made that only the elements produced 
by type II SNe (mostly $\alpha$-elements and some iron) 
can escape the star-forming regions. We have made this choice following the 
conclusions of 
Marconi et al. (1994) and Kunth et al. (1995), 
who showed that models with differential wind 
can explain better the observational constraints of blue compact 
galaxies in general and IZw18 in particular.
The justification for the existence of differential galactic winds
can be found in the fact that during short
starbursts type II SNe dominate.
Since SNII explode in association,
they are likely to produce chimneys which 
will eject metal enriched material (De Young and Gallagher, 1990).
On the other hand, type Ia SNe are not likely to trigger 
a wind since they explode mostly during the interburst phase and have a large
range of explosion times (from $3 \cdot 10^{7}$ to a Hubble time) inducing 
them to explode in isolation. 

The terms on the right side of equation (1) represent, respectively, the
rate at which the gas is lost via astration and galactic wind and the rates at
which the matter is restored to the interstellar medium (ISM) by:\par 
(a)
single stars with masses between $M_L=1.0M_{\odot}$, which is the lowest mass
contributing to the galactic enrichment, and ${M_{B_m}}=3.0M_{\odot}$, which
represents the minimum mass for which a binary system (with the minimum mass 
ratio $\mu_{min}$ defined as in Greggio and Renzini 1983b) produces a type Ia
SN, \par 
(b) binary systems producing type Ia SNe, within the range of masses
${M_{B_m}}$=${3 M_{\odot}}$ and ${M_{B_M}}$=${16 M_{\odot}}$ [the assumed
progenitors of type Ia SNe are binary systems of C-O white dwarfs%
(Whelan and Iben, 1973); the parameter $A$, defined in Greggio and Renzini
(1983b), represents the fraction of binary systems in the IMF
which can give rise to type Ia SNe; 
\par
(c) single stars in the mass range ${M_{B_m}}$ - ${M_{B_M}}$ which end their lives either like white dwarfs or  type II SNe, 
\par
(d) single stars in the mass range ${M_{B_M}}$ and ${M_U}$, where ${M_U}$ is 
the maximum stellar mass contributing to the galactic enrichment 
(100 ${M_{\odot}}$ in our models).
These stars can either end their lives as type II or type Ib SNe. 
\par
The initial mass function (IMF) by mass, $\phi(m)$, 
is expressed as a power law 
with an exponent x=1.35 over the mass range $0.1-100M_{\odot}$. 
\par
The chemical evolution equations include also an accretion term:
$$
\dot G_{iinf}(t)\,=\,{
{(X_{i})_{inf}e^{-t/\tau}}
\over
{\tau(1-e^{-t_{G}/\tau})}
}
$$
where (X$_{i})_{inf}$ is the abundance of the element {\it i} in the infalling
gas, assumed to be primordial, $\tau$ is the time scale of mass accretion 
and $t_{G}$ is the 
galactic lifetime. This accretion term may simulate the formation of
dwarfs as the result of mergers of smaller subunits. 
The parameter $\tau$ has been assumed to be the same for 
all blue compact galaxies and short 
enough to avoid unlikely high infall rates at 
the present time ($\tau=0.5\cdot 10^{9}$ years). 
It is worth noting that a shorter timescale would not produce a noticeable effect on the results.

\subsection{Nucleosynthesis prescription}
\bigskip
The term ${{Q_m}_i(t-{\tau_m})}$ in equation (1), the so called
{\it production matrix} (Talbot and Arnett, 1973), represents the fraction
of mass ejected by a star of mass $m$ in the form of the element $i$. 
The quantity $\tau_{m}$ represents the lifetime of a star of mass m
and $\tau_{m2}$ refers to the mass of the secondary component
and $\tau_{m1}$ to the mass of the primary component of a binary
system giving rise to a type Ia SN (see Matteucci and Greggio, 1986 for
details).
For the
nucleosynthesis prescriptions we have assumed the following:

\smallskip
a) For low and intermediate mass stars (0.8 $\leq$ M/M$_{\odot} \leq$ 
M$_{up}$) we have used Renzini and Voli's (1981) nucleosynthesis 
calculations for a value of the mass loss parameter $\eta$ = 0.33 
(Reimers 1975), and mixing length $\alpha_{RV}$ = 1.5. 
The standard value for 
M$_{up}$ is 8 M$_{\odot}$. 

\smallskip
b) For massive stars (M $>$ 8 M$_{\odot}$) we have used Woosley's (1987) 
nucleosynthesis computations but adopting the relationship between the 
initial mass $M$ and the He-core mass M$_{He}$, from 
Maeder and Meynet (1989). It is worth noting that the adopted M(M$_{He}$) relationship 
does not substantially differ from the original relationship given by  
Arnett (1978) and from the new one by Maeder (1992) based on models with  
overshooting and Z=0.001. These new models show instead 
a very different behaviour of M(M$_{He}$) for stars more massive than 
25~M$_{\odot}$ and Z=0.02, but the galaxies we are modelling never reach 
such a high metallicity.

\smallskip
c) For the explosive nucleosynthesis products, we have adopted the 
prescriptions by Nomoto et al. (1984), model W7, 
for type Ia SNe, which we assume 
to originate from C-O white dwarfs in binary systems (see Marconi 
et al. (1994) for details).
More recent nucleosynthesis calculation by Thielemann et al. (1993) do not
significantly differ from the Nomoto et al. (1984) ones.

\subsection {The nucleosynthesis of nitrogen and the N/O ratio}

Nitrogen is a key element to understand 
the evolution of galaxies with few star forming events since it needs  
relatively long timescales as well as relatively high 
underlying metallicity to be produced. 
The reason is that N is believed to be mostly a  
product of secondary nucleosynthesis,
being produced by CNO processing of $^{12}$C or $^{16}$O  from earlier
generations of stars. 
However, a primary component can be obtained when the 
seed nuclei of $^{12}$C or $^{16}$O are produced in earlier 
helium burning stages of the same star. 
Generally, N is believed to be secondary in massive stars, and mostly 
secondary and probably partly primary in low and intermediate mass stars. 
However, some doubts exist at the moment on the amount of primary 
nitrogen which can be produced in intermediate 
mass stars due to the uncertainties related 
to the occurrence of the third dredge-up in asymptotic giant branch 
stars (AGB). In fact, if Bl\"ocker and Schoenberner (1991) calculations are 
correct, 
the third dredge-up in massive AGB stars should not occur and therefore 
the amount of primary N produced in AGB stars should be 
strongly reduced (Renzini, private communication). 
\par
The only possible way to 
produce a reasonable quantity of N during a short burst (no longer than 
20 Myr), 
as discussed in Kunth et al. (1995),
is to require that massive stars produce a substantial amount 
of primary nitrogen. This claim was already made by Matteucci (1986) in 
order to explain the [N/O] abundances in the solar neighbourhood. 
Recently, calculations by Woosley and Weaver (private communication) 
seem to support the possibility
that N is produced by C and O synthesized inside massive metal poor stars 
as a primary  element.
Therefore, as in Kunth et al. (1995) we have taken this 
possibility into account in the present calculation. 
 
\section{Theoretical prescriptions 
for the evolution of the solar vicinity}

The model adopted for the evolution of the solar neighbourhood is the 
same as in Matteucci and Fran\c cois (1992) and the 
basic equations are similar to Eq (1), the only difference being 
the absence of a galactic wind, namely all the 
$w_{i}=0$.
The main differences between the model for dwarf irregulars and the model for the solar vicinity is that in the latter case the star formation rate is 
continuous and the IMF for massive stars is steeper than a Salpeter IMF
(i.e. Scalo 1986). The time scale for the formation of the disk in
the solar region is also different from 
the timescale assumed for assembling the dwarf 
irregulars (i.e. $\tau=0.5 $Gyr),
namely is $\tau=3$ Gyr. This choice ensures that the majority of the
observational constraints in the solar neighbourhood are reproduced.
The nucleosynthesis prescriptions adopted for the solar neighbourhood are exactly the same as for the dwarfs.

\section{Results}

Besides Model 1 and 5 of Kunth et al. (1995) (model 1 and 3 respectively, in Table 3),
we run several models by varying  star formation efficiency and/or
wind efficiency, all the other parameters being left the same as in Model 5
of Kunth et al. 
The model parameters are presented in Table 3. 
In column 1 are the model 
numbers and in column 2 the nucleosynthesis prescriptions. In particular, 
``STANDARD'' refers to the prescriptions described in section III, whereas 
``N PRIMARY'' refers to the assumption of primary production of N in 
massive stars, leaving all the rest unchanged.

\begin{table}
\caption[]{Model parameters :
~~IMF slope: 1.35 (Salpeter);~~upper mass limit:\\
100 M$_\odot$; lower mass limit: 0.1 M$_\odot$}
\smallskip
\label{}
\begin{tabular}{clllcl}
\hline
 MOD & ~~~~~yields & wind & $\nu$  & $N_{B}$  & duration \\
\smallskip
     &        &      & ($Gyr^{-1}$)    &    & (Myrs) \\
\hline
  1 & STANDARD & 80 & 10 &  1 & 50  \\
       
  2 & N PRIMARY & 700 & 50 & 1  & 20 \\

  3 & N PRIMARY & 80 & 50 & 1 & 20 \\

  4 & N PRIMARY & 700 & 1000 & 1 & 20 \\      
      
  5 & N PRIMARY & 1000 & 50 & 1 & 20 \\
  
  6 & N PRIMARY & 80 & 50 & 4 & 20 \\

  7 & NPRIMARY & 700 & 50 & 4 & 20 \\
\hline
\end{tabular}
\end{table}

In column 3 we show the wind parameter {\it $w_{i}$} 
as defined in the previous 
section. 
This parameter is different from zero only for the elements produced and
dispersed during the explosion of SNe II.
The value for this parameter given in Table 3 refers only
to the $\alpha$-elements, which are the main outcome of 
SN II explosions. In particular, for the elements 
studied here $w_{i}$ is zero for H, He and N, whereas is different 
from zero, but smaller than for $\alpha$-elements, for C and Fe. 
The parameter $w_{i}$ for these elements is chosen in such a way
to account for the fact that Fe and C are produced both in massive and in 
intermediate mass stars (type Ia SNe and AGB stars, respectively).

\par
In column 4 is shown the star formation efficiency $\nu$ (in
units of Gyr$^{-1}$), as defined 
in the previous section. In column 5 is shown the 
number of bursts and in column 6 the duration of each 
burst in Myr. Such a duration is constrained by results of 
population synthesis models (Mas-Hesse and Kunth, 1991) suggesting a
maximum duration for the present burst in IZw18 of 20 Myr. 
The starting time of the burst is not important 
for models 1-5 since we assume
that it is the only event of star formation and it  could happen at any time
(i.e. at any redshift). It is worth noting that
the assumed range of variation of $\nu$ and $w_{i}$ is quite large 
but reasonable. In order to reproduce the star formation rate of IZw18
a value of $\nu$=50 Gyr$^{-1}$ 
(note that for the solar neighbourhood $\nu$=0.5 Gyr$^{-1}$)
is preferred since it predicts a star formation rate
of $\simeq 0.03 M_{\odot}$yr$^{-1}$, in very good agreement 
with observational estimates (see Kunth et al., 1995).
An efficiency of star formation of 1000 Gyr$^{-1}$ is also plausible for starburst 
galaxies since it predicts a star formation rate of 
$\simeq 0.1M_{\odot}$yr$^{-1}$. 
Such star formation rates are quite reasonable for objects suffering
only few bursts of star formation (may be only one) and 
having a large amount of gas.
The wind parameter also spans a quite large range of values and this is possible under the assumption of enriched galactic wind, when only
metals are lost.
On the other hand, in the case of normal wind the value of $w_{i}$
is constrained by the condition of not destroying the galaxy.
However, it is worth noting that even a totally disruptive wind, could in principle be considered.
The initial mass of gas involved in the burst is assumed to be
$6 \cdot 10^{6}M_{\odot}$.

\begin{table}
\caption[]{Measured (O/H) and (N/O) column density ratios
in DLA systems to be compared with the figures. 
}
\smallskip
\label{}
\begin{tabular}{cccr}
\hline
 QSO  & $z_{\rm abs}$ & 12+log(O/H) & log(N/O)~  \\
\smallskip
     &    &    &      \\
\hline
&&& \\
0000-26   & 3.3901  & 5.80  & $-0.52$ \\  
1331+1700 & 1.7765  & 6.12  & $-0.80$ \\  
2348-147 $^a$  & 2.2794  & 6.80  & $< -2.20$ $^a$ \\  
2344+124 $^a$  & 2.5379  & 6.82  & $-2.0$ $^a$ \\  
1946+7658 $^a$ & 2.8443  & 6.28  & $< -2.0$ $^a$ \\ 
&&& \\
\hline
\end{tabular}

\smallskip
$(a)$ Values taken from Lipman (1995).

\end{table}

The N/O versus O/H distribution is shown in Fig. 1
where the models with only one burst together with the models for the solar
neighbourhood are shown.
The dot-dashed curve 
is from model N. 5 from Kunth et al. (1995) (model 3 of table 3)
and it can reproduce the 
observations of IZw 18. 
Incidentally, we note that this model can  fit very well the N/O upper limit
by Pettini et al.  (1995) but for a burst age shorter than 20 Myr.
On the other hand, Pettini's et al.'s point,
as well as the other points from Lipman (1995), could be marginally reproduced
also by the evolution of a disk galaxy in the earliest phases of its 
evolution; in particular during the halo phase, as it is shown in the figure.

\begin{figure}
\vspace{10 cm}
\caption{Log(N/O) vs. 12+log(O/H) as predicted by different models of chemical evolution. In particular, the continuous lines refer to a model
for the solar neighbourhood, as described in the text, for different prescriptions concerning the nucleosynthesis of N. 
The value of the $\alpha$ parameter in Renzini and Voli (1981) is indicated
on each curve; $\alpha$=0 means only secondary N in low and intermediate
mass stars, $\alpha$=1.5 and 2.0 indicate different amounts of primary
N produced during the third dredge-up according to the efficiency of
convection. The line with a plateau and labelled $\alpha$=1.5 represents
a model where N is assumed to be primary in massive stars (in all
the other models is secondary) plus primary N from intermediate mass stars
with $\alpha$=1.5. 
The dash-dotted lines are the predictions of starburst models with only
one burst, 
as described in Table 3.
The symbols refer to measurements of N/O and O/H ratios in the sun 
and in the DLA shown in Table 4. 
In particular, the romb with the highest N/O is the system 
observed by Molaro et al. (1995) whereas the other romb is the DLA system
observed by Green et al. (1995). The clovers refer to different measurements for IZw18, the one with the highest N/O is from Skillman and Kennicutt (1993),
whereas the other is from Dufour et al. (1988). The crossed circles are the 
DLA systems discussed in Pettini et al. (1995) and Lipman (1995).}
\end{figure}

In fact, the continuous curves in Fig. 1 represent the 
predictions from the model of chemical evolution of
the solar neighbourhood under different assumptions 
about the nucleosynthesis of N. In particular,  
the area delimited by those models 
goes from purely secondary N in stars of all
masses (the straight line at the 
right end) to primary N in massive stars and secondary and primary N in
low and intermediate mass stars.
On the other hand, 
the N/O ratios observed by Molaro et al.(1995)  and Green et al.(1995)  
are reproduced by models with 
a strong starburst and very strong galactic wind (models 4 and 5 in Table 2),
so that  
models for the solar neighbourhood seem to be 
completely ruled out in explaining
these two DLA systems.
Thus we show that
it is possible to have 
large N/O ratios even at low O abundances if 
the nitrogen produced by massive stars, restored 
on relatively short time scales, is primary and with a strong differential
effect in the galactic wind. As shown in Fig. 2
the same model produces solar ratios of alpha-elements 
(such as S, O and Si) to iron peak elements (such as Fe and Zn), 
as it is observed at least in few DLA.  
In particular, in Fig. 2 we show the predictions of the models for O and Si
relative to iron to be compared with the data
of Table 4. 
The x-axis does not extend to abundances larger than [Fe/H]=-2.0
since in one-burst models the metallicity does not increase any further.
Sulphur is not shown here but it should closely
follow oxygen.
On the other hand, the predicted [Si/Fe] is lower than [O/Fe]
and the reason for this resides in the fact that 
more Si than O is produced in type Ia SNe
(Nomoto et al. 1984; Thielemann et al. 1993) and that we assume
the same $w_{i}$ parameter for Si and O.
Studies of abundances in halo stars do not allow us
to discriminate clearly on this point
(Fran\c cois, 1986; Ryan et al. 1991; Primas et al. 1994;
McWilliam et al. 1995). 
Therefore, it does not seem safe 
to assume [Si/O]=0 in order to derive oxygen abundances in DLA systems
(see Lipman, 1995).
However, as indicated in Figure 2 the observed [O/Fe] is lower than [Si/Fe]
for the DLA observed by Molaro et al. (1995). A possible explanation for this
is that,
in the framework of the differential galactic wind, O,
which originates mostly in type II SNe, should be lost from
the galaxy in a larger percentage than silicon, which  
originates also from type Ia SNe. 
In other words, Si should be treated as we do with  C and Fe.
However, numerical experiments show that even in this case is difficult to 
invert the situation and have [Si/Fe] higher than [O/Fe], as indicated by the
data taken at face values!

\begin{figure}
\vspace{10 cm}
\caption{ Predicted and observed [O,Si/Fe] versus [Fe/H]. The 
long-dashed lines refers to O, whereas the short-dashed lines refers to Si.
The model numbers are indicated on each curve. 
}
\end{figure}

\begin{figure}
\vspace{10 cm}
\caption{The same as figure 1 but here are shown only the predictions of models 6 (dashed-dotted line) and 7 (dotted line).}
\end{figure}

In Figure 3 we show the predicted N/O vs. O/H from models with
4 short bursts (20 Myr each) all occurring inside the first Gyr from 
the start of star formation, so that they can be representative of high redshift objects. In particular, we show the predictions of model 6 and 7 which are
similar to model 3 and 2, respectively, but with four bursts of star formation.These models show that in principle the DLA systems from Green et al. (1995)
and from Molaro et al. (1995)
could be explained by a galaxy like IZw18 
experiencing
more than one burst of star formation and observed during 
the interburst period, without invoking an
extremely large wind parameter. In fact, during this phase the N/O ratio 
increases as due to the fact that oxygen is no more produced while N
continues to be restored from low and intermediate mass stars.
This is a well known effect as shown by Pilyugin (1992;1993).

In Figs. 4 and 5 we show the [O,Si/Fe] ratios predicted by model 6 and 7,
respectively. 
Here too we can see the oscillating behaviour due to the alternating burst and quiescent phases. In particular, in both models the [O,Si/Fe] ratios
decrease during the interburst phase and increase again during the bursts.
The difference between the two models resides in the efficiency of 
the wind in model 7, which reflects in a stronger variation of the abundance
ratios between the burst and the interburst phases. 
The reason for this variation in model 7 is that the stronger 
wind acting during bursts is responsible for very low absolute 
abundances at the end of each burst, so 
the increase of Fe during the interburst is stronger relatively
to model 6 where the absolute abundances at the end of each burst are higher. 
\par
Finally,
it is worth noting that current models for elliptical galaxies
(see Matteucci and Padovani, 1993) would never reproduce such the high 
N/O ratio observed by Molaro et al. (1995)
for such low metallicities. 
In Figure 1, for example, the predictions 
of models for an elliptical galaxy of initial luminous
mass $10^{11} M_{\odot}$ would lie at the right side of the solar 
neighbourhood curve for only secondary nitrogen, 
and even in the case of primary N from massive stars 
would be completely outside of the 
metallicity region where the DLA systems are observed.

\begin{figure}
\vspace{10 cm}
\caption{The same as figure 2 but for model 6. The dashed 
lines refer to O, whereas the dotted lines refer to Si.}
\end{figure}

\begin{figure}
\vspace{10 cm}
\caption{The same as figure 2 but for model 7. The dashed 
lines refer to O, whereas the dotted lines refer to Si.}
\end{figure} 

\section{Conclusions}

The abundance pattern of non refractory elements
in some primeval DLA systems at high redshift
seems not to follow the pattern observed in the halo of the Milky
Way. This suggests that these objects 
had a different chemical
evolution than the Milky Way and the spirals in general, at variance with
the general belief that DLA are the progenitors of present day spirals.
We have shown that
the most promising models to explain the observed abundances
are those succesfully applied to the dwarf irregular galaxies such as
IZw18 (Marconi et al. 1994, Kunth et al. 1995).  
The conclusion about these DLA systems being dwarf galaxies
has also  been independently suggested by 
Meyer and York (1992) and by  Steidel et al.
(1994) from the low abundances found in the few
DLA observed at low redshifts.
\smallskip

In summary, our conclusions are:\par
- In order to explain the high N/O ratios observed in two DLA systems
by Green et al .(1995) and Molaro et al.(1995) one has to assume that these systems are dwarf irregular galaxies experiencing their first or one of their first bursts of star formation. 
These galaxies should also experience strong enriched galactic winds carrying away mostly the products of SN II explosions such as oxygen and other 
$\alpha$-elements. 
In particular, the high N/O abundance ratios could represent
either the situation of an interburst phase
where N increases and O does not,
or the situation of a burst triggering an extremely strong and enriched 
galactic wind.
\smallskip

-Nitrogen in massive stars should have a primary origin as 
already suggested by Matteucci (1986). This seems to be possible but 
is strongly dependent on the
assumed treatment of convection in stellar interiors.
The nucleosynthesis of N in stars of all masses, especially in low and
intermediate mass stars, which are the main producers of this element, needs
revision and a homogeneous set of calculations for stars of 
all masses is required. New yields, but only for stars below 4 $M_{\sun}$,
have been recently computed by Marigo et al. (1996). Unfortunately,
the mass range above that mass limit is very important for nitrogen
production and therefore firm conclusions cannot yet be drawn. 

\smallskip

-We can exclude, on the basis of current models for elliptical galaxies that
any of the systems discussed in this paper could be a proto-elliptical.

\smallskip

- Some of the differences in the abundances observed among different 
DLA systems could be due to the fact that some of them are proto-spirals
(see the systems observed by Pettini et al. 1995 and Lipman, 1995) 
and some are proto-dwarfs (see the systems observed by Molaro et al. 1995 and
Green et al. 1995)

\smallskip
-  Abundance ratios between elements produced from stars at different rates
such as N/O and $\alpha$/Fe
represent a very useful tool either to date a galaxy
or to understand the nature
of high red-shift objects.


\begin{thebibliography}{}

\bibitem[]{}
Anders E., Grevesse N., 1989, Geochim. Cosmochim. Acta 53, 197

\bibitem[]{}
Arnett D.W., 1978, ApJ 219, 1008

\bibitem[]{}
Black J.H., Chaffee F.H., Foltz C.B., 1987, ApJ 317, 442

\bibitem[]{}
Bl\"ocker T., Schoenberner D., 1991, A\&A 244, L43

\bibitem[]{}
Chaffee F.H., Black J.H., Foltz C.B., 1988, ApJ 335, 584

\bibitem[]{}
De Young D.S., Gallagher J.S. III, 1990, ApJ 356, L15

\bibitem[]{}
Dufour R.J., Garnett D.R., Shields G.A., 1988,  ApJ 332, 752

\bibitem[]{}
Fall S.M., Pei Y.C., 1989, ApJ 337, 7


\bibitem[1994]{ft}
Fan X.-M., Tytler D., 1994, ApJ Suppl. 94, 17

\bibitem[]{}
Fran\c cois P., 1986, A\&A 160, 264

\bibitem[]{} 
Gerola H., Seiden P.E., Schulman L.S., 1980, Ap.J., 242, 517


\bibitem[]{}
Green R.F.,  York D., Huang  K.,  Bechtold J., Welty D.,
Carlson M., Khare P., Kulkarni V., 1995, Proc. 
{\it ESO Workshop on QSO Absorption Lines}, ed. G. Meylan,
Springer Verlag, 85


\bibitem[]{}
Greggio L., Renzini A., 1983a, in {\it The first stellar generations},
Mem. S.A.It. Vol. 54, 311

\bibitem[]{}
Greggio L., Renzini A., 1983b, A\&A 118, 217

\bibitem[]{}
Hamman F., Ferland G., 1993, ApJ, 418, 11 

\bibitem[]{}
Hannaford P., Lowe R.M., Grevesse N., Noels A., 1992, A\&A 259, 301

\bibitem[]{}
Kulkarni V.,P., Huang K., Green R.F., Bechtold J., Welty D., York D.G., 
1996, MNRAS 279, 197

\bibitem[]{}
Kunth D., Matteucci F., Marconi G., 1995, A\&A 297, 634

\bibitem[]{} 
Lequeux J., Kunth D., Mas-Hesse J.M., Sargent W.L.W., 1995, A\&A 301, 18


\bibitem[]{}
Lipman K., 1995, PhD thesis, Cambridge University 

\bibitem[1995]{}
Lu L., Savage B.D., Tripp T.M., Meyer D.M. 1995a, ApJ 447, 597

\bibitem[]{}
Lu L., Sargent, W.L.W., Womble, D.S.,Barlow, T.A. 1995b, ApJ in press

\bibitem[]{} 
Maeder A., Meynet G., 1989, A\&A, 210, 155

\bibitem[]{} 
Maeder A., 1992, A\&A, 264, 105

\bibitem[]{}
Marconi G., Matteucci F., Tosi M., 1994, MNRAS 270, 35

\bibitem[]{}
Marigo P., Bressan A., Chiosi C., 1996, preprint

\bibitem[]{}
Mas-Hesse J.M., Kunth D., 1991, A\&A Suppl. Ser. 88, 399

\bibitem[]{}
Matteucci F., 1995, in {\it The interplay between massive star formation,
ISM and galaxy evolution}, Eds. Kunth et al., Edition Frontiers, in press

\bibitem[]{}
Matteucci F., 1986, MNRAS, 221, 911




\bibitem[]{}
Matteucci F., Fran\c cois, P. 1992, A\&A, 262, L1 

\bibitem[]{}
Matteucci F., Greggio L., 1986, A\&A, 154, 279 

\bibitem[]{} 
Matteucci F., Padovani P., 1993, Ap. J., 419, 485

\bibitem[]{} 
Matteucci F., Tosi M., 1985, MNRAS., 217, 391

\bibitem[]{}
Matteucci F., Raiteri C.M., Busso M., Gallino R., Gratton R.,
1993, A\&A, 272, 421

\bibitem[]{}
McWilliam A., Preston G., Sneden C., Searle L., 1995, Astron J. 109, 2757

\bibitem[]{} 
Meurer G.R., Freeman K.C., Dopita M.A., 1992, Astron. J. 103, 60


\bibitem[1990]{}
Meyer D.M., Roth K.C., 1990, ApJ 363, 57

\bibitem[1989]{}
Meyer D.M., Welty D.E., York D.G., 1989, ApJ 343 L37

\bibitem[1992]{}
Meyer D.M., York D.G., 1987, ApJ 319, L45

\bibitem[1992]{}
Meyer D.M., York D.G., 1992, ApJ 399, L121

\bibitem[]{}
Mobasher, B., Rowan-Robinson, M., Georgakakis, A., Eaton, N., 1996, 
MNRAS, in press



\bibitem[]{}
Molaro P., D' Odorico S., Fontana A., Savaglio S., Vladilo G.,
1995, A\&A in press (ESO sc. prepr. No. 1102)

\bibitem[]{} 
Nomoto K., Thielemann F.K., Yokoi, K., 1984, Ap.J., 286, 644

\bibitem[]{} 
Papaderos P., Fricke K., Thuan T.X., Loose H., 1994, A\&A, 291, L13

\bibitem[]{}
Pettini M., Boksenberg A., Hunstead R.W., 1990, ApJ, 348, 48

\bibitem[]{}
Pettini M., Lipman K., Hunstead R.W., 1995, ApJ 451, 100

\bibitem[1994]{pshk}
Pettini M., Smith L.J., Hunstead R.W., King D.L.,
1994, ApJ 426, 79

\bibitem[]{} Pilyugin, L.S., 1992, A\&A, 260, 58

\bibitem[]{} Pilyugin, L.S., 1993, A\&A, 277, 42

\bibitem[]{} Primas, F., Molaro, P., Castelli, F., 1994, A\&A, 290, 885


\bibitem[]{}
Rauch M., Carswell R.F., Robertson J.G., Shaver P.A., Webb J.K.,
1990, MNRAS 242, 698

\bibitem[]{} 
Reimers D., 1975, M\'em. R. Sci. Li\`ege 6\`eme S\'er., 8, 369

\bibitem[]{} 
Renzini A., Voli M., 1981, A\&A, 94, 175

\bibitem[]{}
Ryan S.G., Norris J.E., Bessel M.S., 1991, Astron. J. 102, 303

\bibitem[]{}
Scalo J.M., 1986, Fund. Cosmic Phys. 11, 1

\bibitem[]{}
Skillman E.D., Kennicutt R.C., 1993, ApJ 411, 655

\bibitem[]{}
Steidel C.C., Bowen D.V., Blades J.C., Dickinson M., 
1995, ApJ 440, L45

\bibitem[]{}
Steigman, G., Strittmatter, P.A., Williams, R.E. 1975, ApJ, 198, 575

\bibitem[]{}
Talbot R.J., Arnett D.W., 1973, ApJ 197, 551

\bibitem[]{}
Tinsley B.M., 1980, Fund. Cosmic Phys. 5, 287

\bibitem[]{}
Thielemann F.K., Nomoto K., Ashimoto M., 1993, in {\it Origin
and evolution of the elements}, Eds. N. Prantzos et al.,
Cambridge Univ. Press, 297
 
\bibitem[1995]{}
Vladilo G., D'Odorico S., Molaro P., Savaglio S., 1995,
Proc. {\it ESO Workshop on QSO absorption lines}, 
ed. G. Meylan, Springer Verlag, 103

 
\bibitem[]{}
Whelan J., Iben I. Jr., 1973, ApJ 186, 1007

\bibitem[1994]{wtsc}
Wolfe A.M., Turnshek D.A., 
Smith H.E., Cohen R. D., 1986, ApJS 61, 249

\bibitem[]{} 
Woosley S.E., 1987, in ``Nucleosynthesis and Chemical Evolution''
  16th Saas-Fee Advanced Course, Geneva Observatory, p.1

\end{thebibliography}
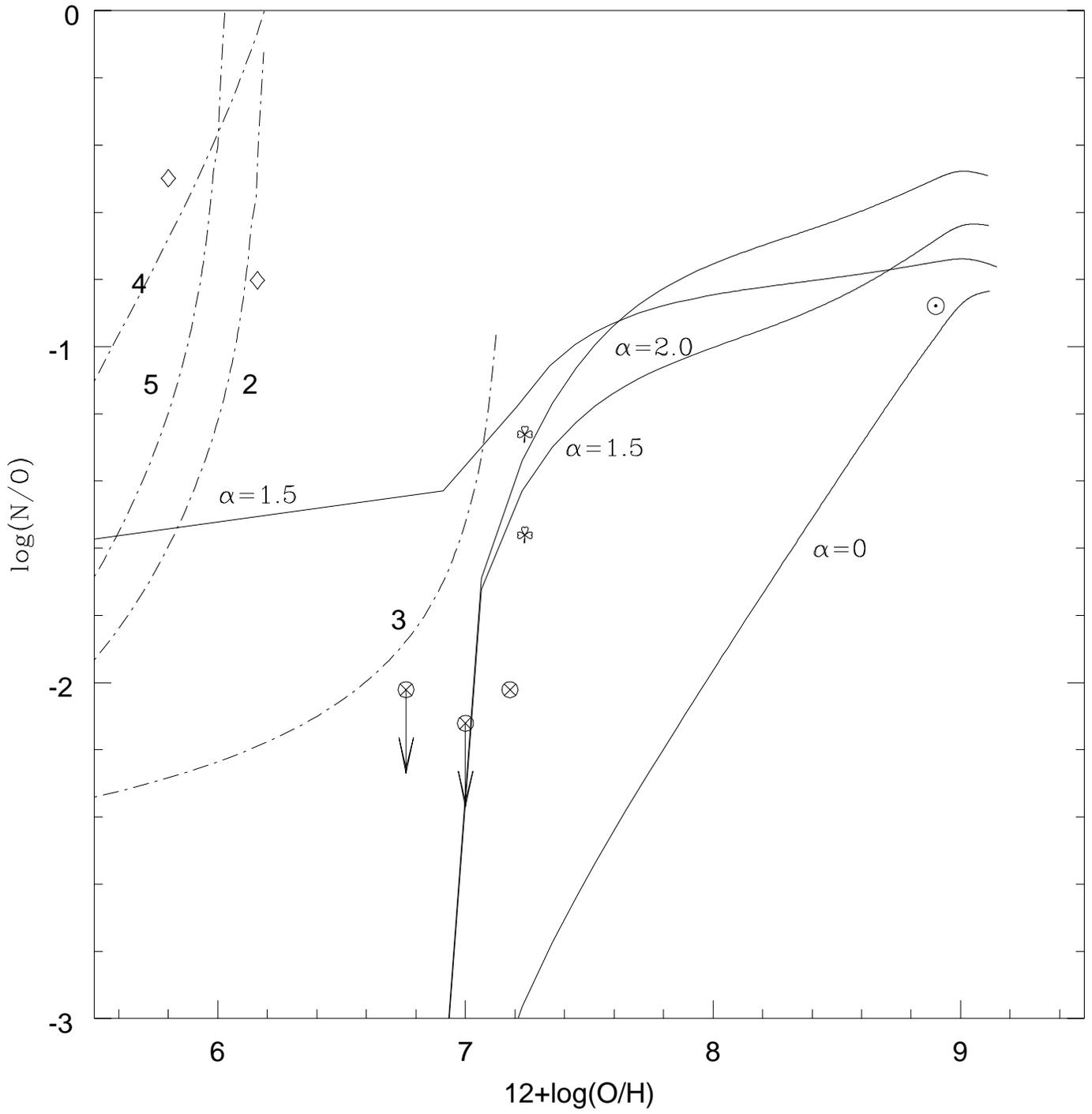
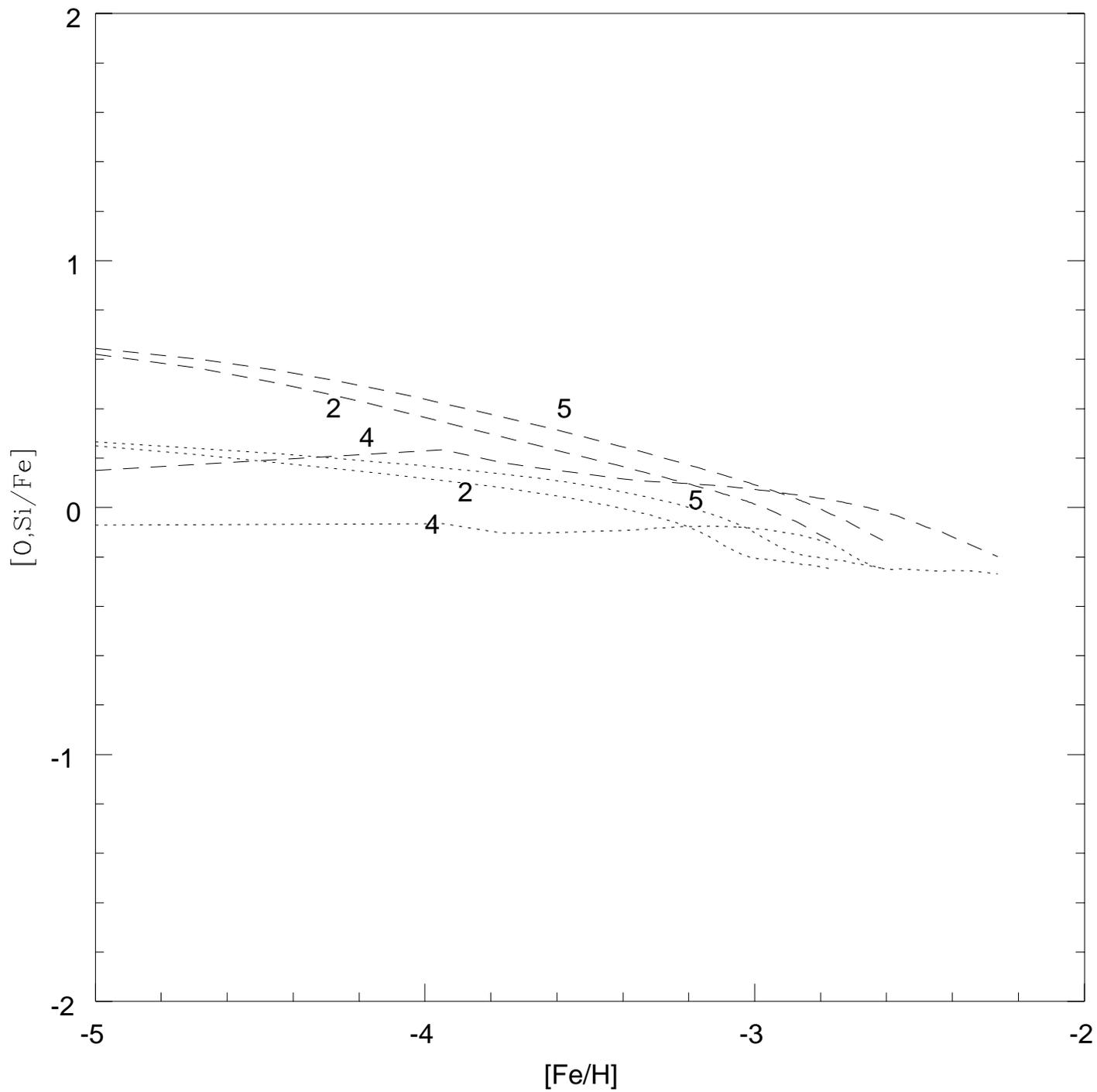
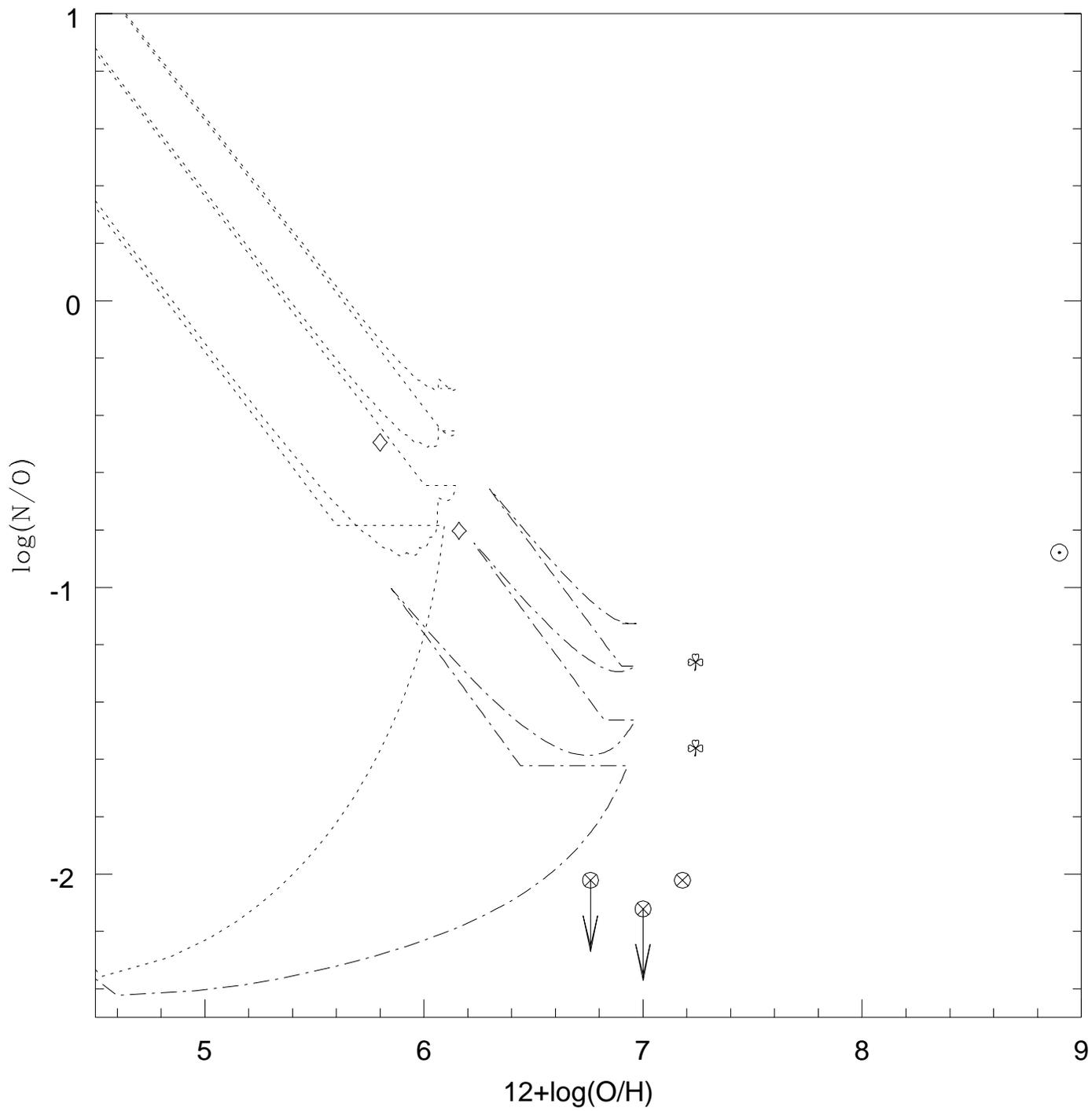
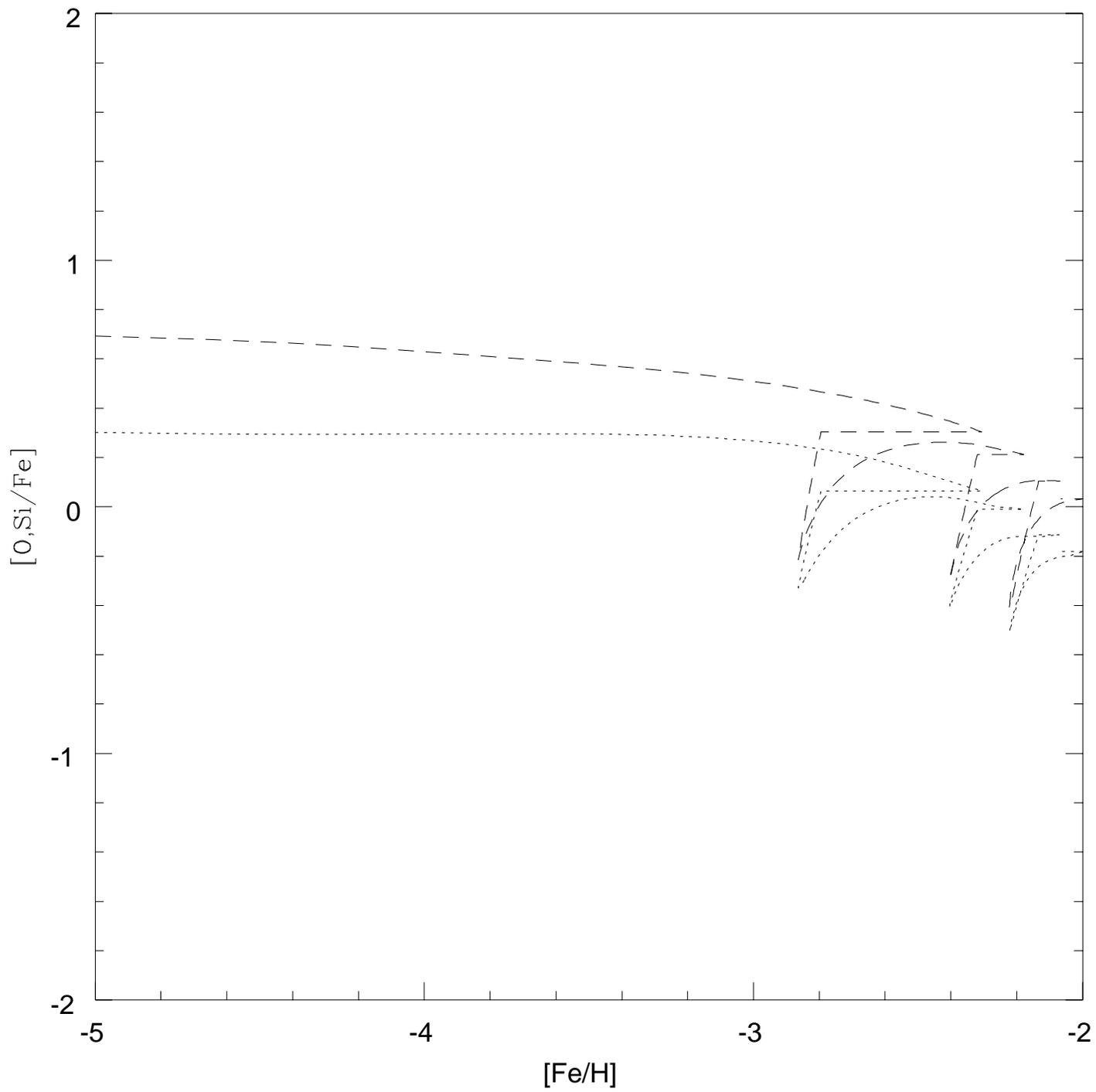
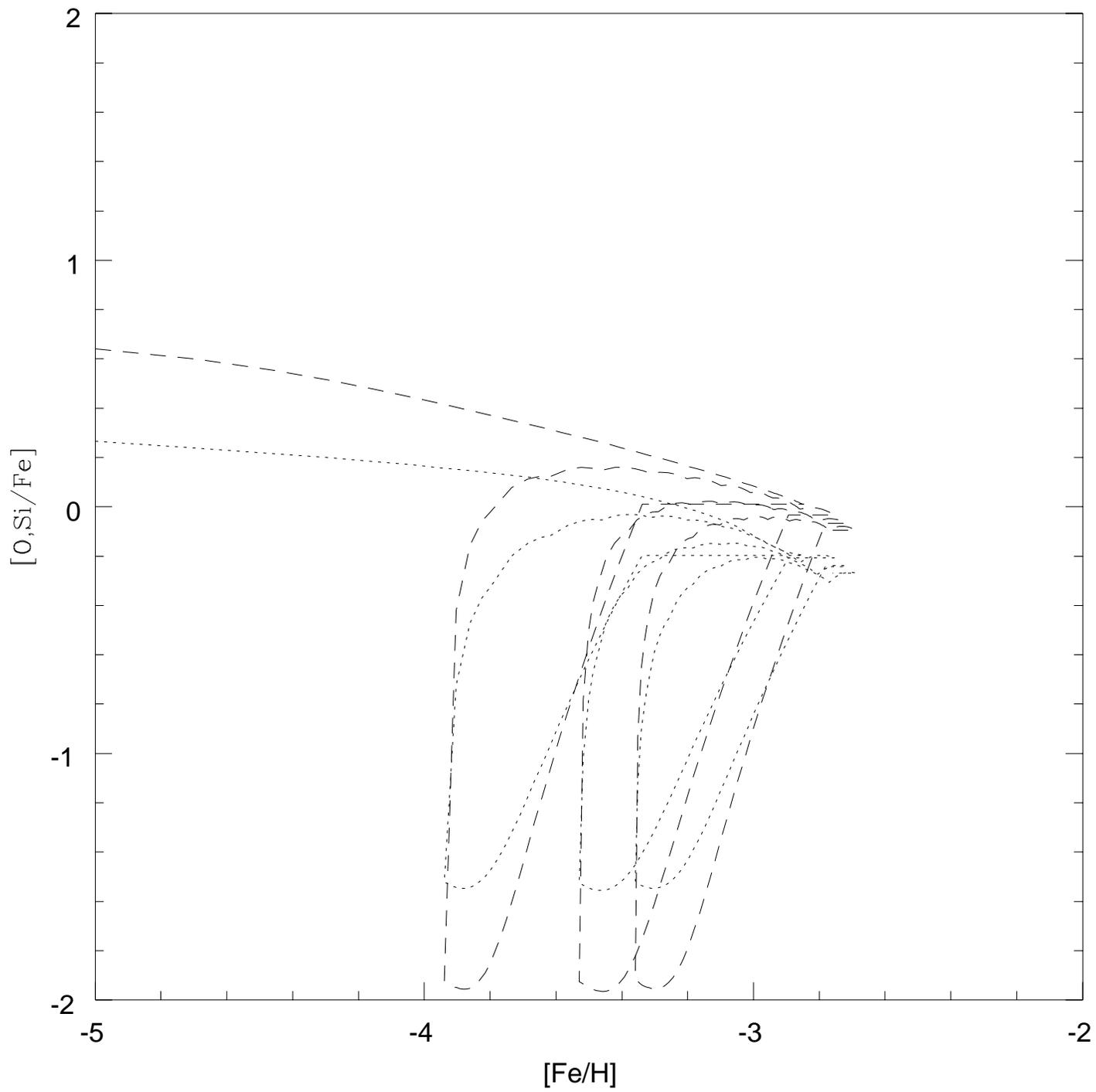
\end{document}